\documentclass[12pt]{iopart}

\newcommand{\ket}[1]{\left|#1\right\rangle}
\usepackage{graphicx}
\usepackage{amssymb}

\begin{document}

\title[Distributed cavity phase frequency shifts of the caesium fountain PTB-CSF2]{Distributed cavity phase frequency shifts of the caesium fountain PTB-CSF2}
\author{S. Weyers$^1$, V. Gerginov$^1$, N. Nemitz$^1$, R. Li$^2$ and K. Gibble$^2$}
\address{$^1$ Physikalisch-Technische Bundesanstalt (PTB), Bundesallee 100, 38116 Braunschweig, Germany\\
$^2$ Department of Physics, The Pennsylvania State University, University Park, PA 16802, USA}
\ead{stefan.weyers@ptb.de}

\begin{abstract}

We evaluate the frequency error from distributed cavity phase in the caesium fountain clock PTB-CSF2 at the Physikalisch-Technische Bundesanstalt with a combination of frequency measurements and {\it ab initio} calculations. The associated uncertainty is $1.3 \times 10^{-16}$, with a frequency bias of $0.4 \times 10^{-16}$. The agreement between the measurements and calculations explains the previously observed frequency shifts at elevated microwave amplitude. We also evaluate the frequency bias and uncertainty due to the microwave lensing of the atomic wavepackets. We report a total PTB-CSF2 systematic uncertainty of $4.1 \times 10^{-16}$.

\end{abstract}

\maketitle

\section{\label{sec:introduction}Introduction}

Over the past decade the number of primary caesium fountain clocks contributing to International Atomic Time (TAI) has steadily increased and now exceeds the number of contributing primary caesium beam clocks. During the last two years, ten fountain clocks from seven laboratories contributed to TAI~\cite{BIPMCircularT}. The fountain clocks largely calibrate TAI because of their excellent systematic and statistical uncertainties. While total systematic uncertainties clearly below $10^{-15}$ are routinely reported by several fountains, there has not been widespread agreement about the methods to establish the frequency uncertainties due to spatially distributed phase variations of the field in the microwave cavity.

Until recently, distributed cavity phase (DCP) was the largest uncertainty contribution for several fountains~\cite{Wynands2005,Li2010,Guena2011,Li2011}. The resistive losses of the cavity walls lead to small traveling waves, which account for the power flow from the cavity feeds to the walls and produce the spatial phase variations~\cite{Li2004}. A recent comprehensive model has clarified the evaluation of DCP frequency shifts~\cite{Li2010}. The {\it ab initio} model, with no free parameters, has been stringently verified with the French fountain SYRTE-FO2 and already used to evaluate the DCP uncertainties of it and the British NPL-CsF2 fountain~\cite{Guena2011,Li2011}. Here we report the results of measurements and calculations, using this model to evaluate the DCP uncertainty of the PTB-CSF2 fountain clock. In contrast to SYRTE-FO2 and NPL-CsF2, the amplitudes and phases of the two cavity feeds of PTB-CSF2 are balanced by symmetry and cannot be independently adjusted~\cite{Gerginov2010,Schroeder2002}. Here, we also evaluate the microwave lensing frequency shift~\cite{Li2011,Gibble2006} and summarize all of the evaluated systematic biases and uncertainties of PTB-CSF2.

The fountain geometry naturally minimizes frequency shifts from phase variations of the microwave field. In atomic fountain clocks, the atoms interact twice with the microwave field in the same cavity, eliminating the end-to-end phase shifts of beam clocks~\cite{Vanier1989,Bauch1998} with a velocity reversal, which largely cancels the Doppler shifts from longitudinal phase gradients. Transverse phase variations clearly produce frequency shifts when the atoms probe the field with different phases when they traverse the cavity at different positions during their ascent and descent in a fountain.

The central feature of the model~\cite{Li2004,Li2010} is to decompose the transverse phase variations into a Fourier series $\cos(m \phi)$ with cylindrical coordinates $(\rho,\phi,z)$. This enables densely-meshed finite element calculations of the cavity fields because the 3D problem is reduced to a short series of 2D solutions in $\rho$ and $z$. The transverse phase of the field is proportional to $\rho^m \cos(m\phi)$ to lowest order and therefore only $m \le 2$ terms are significant since the cavity apertures restrict the atomic trajectories to small $\rho$. The azimuthal symmetry of each Fourier term specifies the physical effects in the fountain that produce DCP frequency shifts. While the $m=0$ transverse variations are small, the $m=1$ phase variations are naturally large, representing a linear phase gradient and a power flow from a cavity feed across the cavity to the walls. This gradient is highly suppressed in all fountains by using two or more diametrically opposed feeds. The quadrupolar $m=2$ transverse phase variations are maximally excited by two opposing feeds and can be negligible with four (or more) azimuthally distributed feeds~\cite{Jefferts2002}.

The DCP shifts of the lowest Fourier terms usually have unique functional dependences on the clock's microwave amplitude. For example, if the transverse phase variations have no longitudinal variation for all atomic trajectories, then each atom sees the same phase for the entire cavity traversal during ascent and descent. Then, the DCP frequency shifts are independent of microwave amplitude. But, in general, there are longitudinal phase variations and therefore the atoms see a time-dependent phase during each cavity traversal, which often gives a large microwave amplitude dependence of the effective phase of the cavity field~\cite{Li2010,Li2005}. Even longitudinal phase gradients that have no transverse phase variation can produce large amplitude-dependent frequency shifts because the atoms do not have a transverse velocity reversal in the fountain and therefore experience different average microwave pulse areas on the two cavity passages~\cite{Li2010,Li2005}. The $m=0$, azimuthally symmetric phase deviations have a negligibly small transverse variation and a very large longitudinal variation, primarily due to the endcap losses~\cite{Li2004}. Here we measure the resulting large, $m=0$ amplitude-dependent frequency shift, explaining the previously observed frequency shifts of the PTB-CSF1 and PTB-CSF2 fountains at elevated microwave amplitudes~\cite{Weyers2007,Gerginov2010EFTF}.

\section{\label{sec:Evaluation}DCP evaluation of PTB-CSF2}

The cylindrical TE$_{011}$ Ramsey cavity of PTB-CSF2 is made from OFHC copper, with an inner diameter of 48.4\,mm and 28.5\,mm height~\cite{Gerginov2010,Schroeder2002}. In the top and bottom endcaps two apertures 10~mm in diameter enable the atoms to pass through the cavity, with 70\,mm long waveguide cutoff tubes to minimize microwave leakage from the cavity. The coaxial line that feeds the microwave signal is strongly coupled to a standing wave in a curved waveguide. Two opposed slits (10.8\,mm length and 2\,mm width) in the curved waveguide feed the cavity. Their strong coupling to the cavity gives a relatively low loaded quality factor $Q=1600$, which reduces the effect of the cavity detuning, for example due to temperature changes, as compared to weak coupling. Our finite-element calculations~\cite{Li2004,Li2010} of the cavity geometry give consistent resonant frequencies for the measured TE$_{011}$ and TM$_{111}$ modes.

In PTB-CSF2 the entire atom collection and cooling chamber, including the laser beam collimators, can be pivoted around its center on a goniometric stage. It allows a sub-milliradian adjustment of the launch direction to maximize the number of detected atoms. The initial center-of-mass position of the atomic cloud is offset from the fountain axis by 1.7\,mm, along the axis of the two cavity feeds. This is inferred from the single Rabi pulse spectrum of $\ket{F=3, m_F=0}\rightarrow \ket{F=4, m_F=\pm 1}$ transitions, taking advantage of a slight unintended tilt of the static magnetic field in the vertical plane perpendicular to the feeds~\cite{Nemitz2011}. Because of this offset, and since PTB-CSF2 is vertically adjusted to maximize the number of detected atoms, the actual angle $\alpha_{\|}$ at which the detected atoms are launched deviates from true vertical by 0.15\,mrad so that the atomic cloud is launched with a nonzero horizontal launch velocity of 0.7\,mm/s in opposite direction of the offset. Perpendicular to the feed axis, the above evaluation of the spectrum cannot be used since the static magnetic field happens to have a negligible component parallel to the feeds~\cite{Nemitz2011}. Instead we evaluate the $\ket{3,0}\rightarrow \ket{4,1}$ transition probabilities for a single Rabi pulse on the ascent as a function of the launch angle perpendicular to the feed axis. This transition probability shows no gradient at the launch angle that maximizes the number of detected atoms for a cloud that is launched on the cavity axis. We observe no gradient of the transition probability versus perpendicular tilt and therefore conclude that there is no cloud offset perpendicular to the feed axis. For tilts parallel to the feed axis, this technique is consistent with the above measured offset because there is a gradient of the $\ket{3,0}\rightarrow \ket{4,1}$ transition probability versus launch angle.

\begin{figure}
\includegraphics[width=15cm]{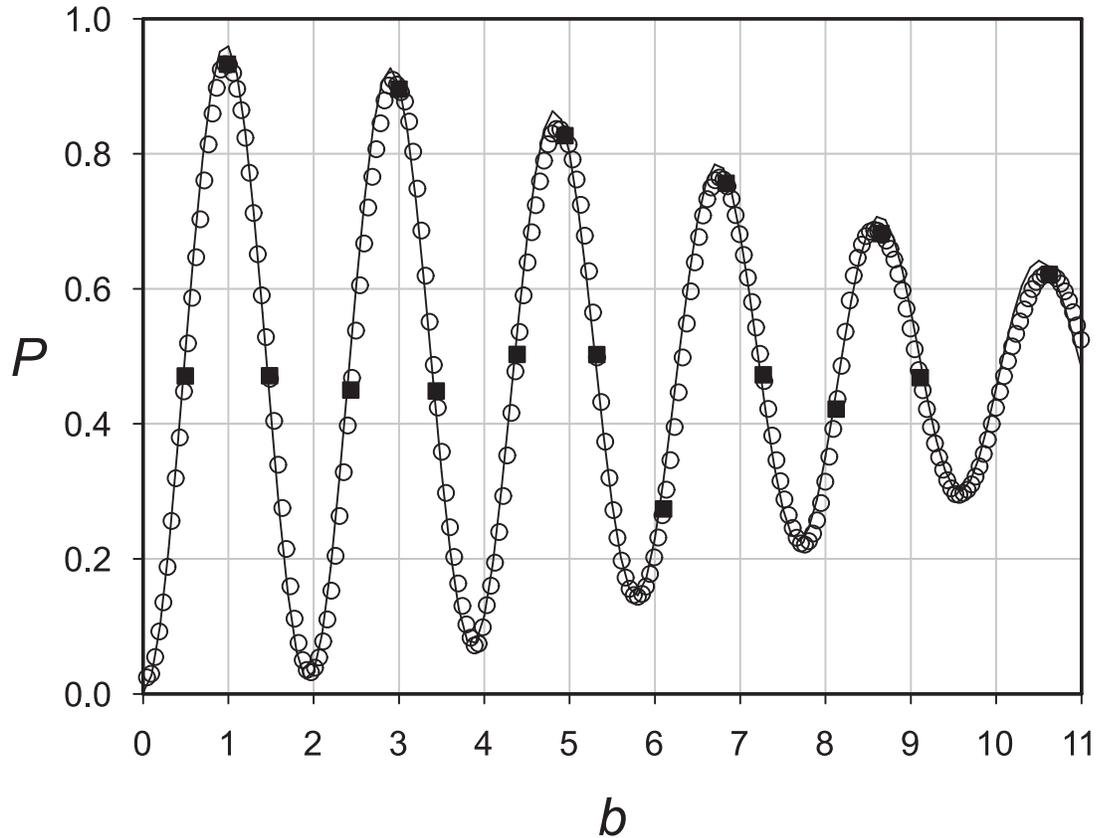}
\caption{\label{fig:transitionP} Transition probability $P$ in PTB-CSF2 versus microwave amplitude $b$. The open circles indicate measured transition probabilities. The squares indicate the amplitudes at which the $m=0$ DCP contribution was measured (see figure~\ref{fig:m=0} below) and the solid line is from the model.}
\end{figure}

The validity of the Monte-Carlo model of the fountain is confirmed by comparing experimental and calculated transition probabilities $P$ as a function of microwave amplitude $b$ (see figure~\ref{fig:transitionP})~\cite{Guena2011,Guena2011b}. The agreement is sensitive to the size, offset, and velocity distributions, the positions of apertures in the fountain, and the detection inhomogeneity. Here $b$ is in units of $\pi/2$-pulses during a cavity traversal for a uniform density distribution on a cavity passage~\cite{Li2010}. Because the cloud size in PTB-CSF2 is smaller than the cavity aperture during the first cavity passage, the optimum microwave amplitude for normal clock operation, which maximizes the Ramsey fringe contrast, is $b=0.97$.

We next evaluate the DCP frequency shifts from each Fourier component $m=0,1,2$. The frequency shifts are often singular near even $b$ where the Ramsey fringe contrast vanishes. Therefore, we instead discuss the change in transition probability $\delta P$ at detuning $+\Delta \nu/2$, where $\Delta \nu=0.912$\,Hz is the full width of the central Ramsey fringe. The frequency shift is $\delta P / (dP/d\nu)$ where the Ramsey fringe slope for PTB-CSF2 is $dP/d\nu=1.57/$Hz at optimum amplitude.

\subsection{\label{sec:m=0}$m=0$ DCP frequency shift}

The $m=0$ DCP shift is primarily due to power flow from the feeds at the cavity midplane to the endcaps. This power flow causes large longitudinal phase variations, which are symmetric about the cavity midplane, and small, azimuthally symmetric transverse phase variations. The longitudinal phase variations are not canceled by the velocity reversal of the two cavity passages because the cloud is larger for the descent than the ascent, and therefore experiences a different average field amplitude~\cite{Li2010,Li2005}. This gives large frequency shifts at elevated microwave amplitude, especially $b=4$ and 8 where the sensitivity function gives the maximum perturbation for a phase variation that is symmetric about the cavity midplane. At low amplitude, $b\ll 1$, $\delta P$ is small because the effective phase is zero since it is simply the average phase along a trajectory through the cavity. At $b = 2$, $\delta P$ is also zero because the phase is an even function of $z$~\cite{Li2010,Li2005}. Thus, $\delta P$ is highly suppressed at optimum amplitude because, while it increases from zero as $b^4$ for small $b$, it cannot get too large since it returns smoothly to zero near $b=2$.

\begin{figure}
\includegraphics[width=15cm]{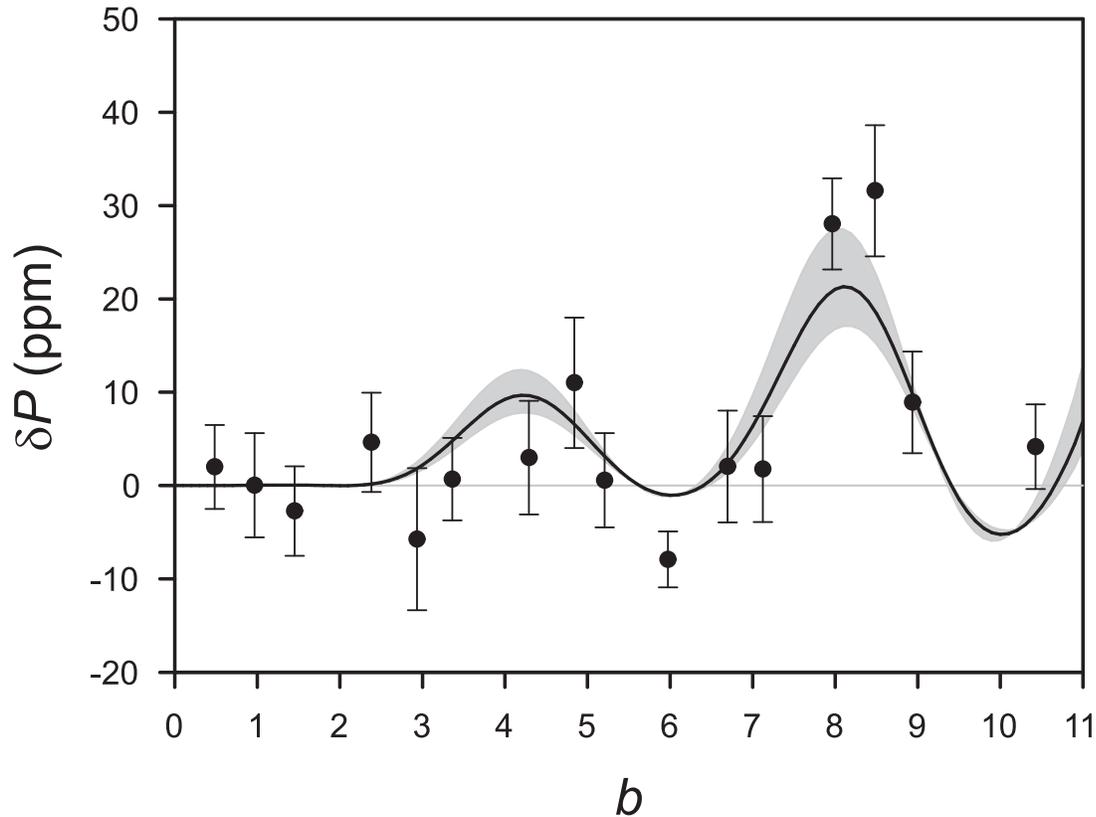}
\caption{\label{fig:m=0} The $m=0$ DCP change of transition probability $\delta P$ in PTB-CSF2 as a function of microwave amplitude $b$. The initial cloud offset is 1.7\,mm. Solid line: Calculated $\delta P$ for an initial cloud radius at launch of 3.9\,mm ($1/\sqrt{e}$). Gray band: Range of shifts for initial cloud sizes of $\pm$20\%.  Full circles: Measurements.}
\end{figure}

The solid curve in figure~\ref{fig:m=0} shows the calculated changes $\delta P$ of transition probability as a function of microwave amplitude for PTB-CSF2. The initial measured $1/\sqrt{e}$ cloud radius at launch is 3.9\,mm and the temperature is 1.2\,$\mu$K. If the cloud is initially centered (no offset), the $\delta P$ is almost identical. The curves show the characteristic features of $m=0$ DCP shifts: a very small shift around optimum amplitude $b\lesssim 1$ and large shifts near $b = 4$ and $b = 8$. At $b=8$ the $m=0$ DCP shift is $7.9 \times 10^{-15}$.

We extract the experimental change in transition probability in figure~\ref{fig:m=0} (full circles) from measurements of the frequency difference between the PTB-CSF2 and PTB-CSF1 fountains, where PTB-CSF1~\cite{Weyers2001} is operated at optimum amplitude. We also measure the Ramsey fringe slopes of PTB-CSF2 at each $b$. We apply frequency corrections to both fountains for the quadratic Zeeman effect, cold collisions, blackbody radiation, gravity and the relativistic Doppler shift. The error bars are 1\,$\sigma$ statistical uncertainties. At optimum amplitude of PTB-CSF2, $b=0.97$, the two fountains agree within the statistical uncertainty. 

The measurements in figure~\ref{fig:m=0} must also include $m=1$ and $m=2$ DCP shifts. Below we show that both are negligible on this scale (see sections~\ref{sec:m=1} and~\ref{sec:m=2}). The calculated $m=0$ DCP shifts agree well with the measurements, which explains the previously observed amplitude dependence of the PTB-CSF2 fountain~\cite{Gerginov2010}.

At optimum amplitude the calculated frequency shift is $1.8 \times 10^{-18}$. Because the  $m=0$ DCP shift depends strongly on the difference of the average microwave amplitudes that the atoms experience on the two cavity passages, the shift is sensitive to the initial cloud size. In figure~\ref{fig:m=0} the gray band shows the range of shifts for initial cloud sizes of $\pm$20\%. This range of initial cloud size gives a negligible uncertainty of $6 \times 10^{-19}$. In~\cite{Guena2011}, measured $m=1$ DCP shifts suggested that the cavity walls could have inhomogeneous conductivities as large as 20\%. If the conductivity of the top and bottom endcaps differ by 20\%, it would give an $m=0$ DCP shift of $3.8 \times 10^{-18}$ at optimum amplitude. The measurements in figure~\ref{fig:m=0} do not exclude this possibility and we therefore take this as our $m=0$ DCP uncertainty.

\subsection{\label{sec:m=1}$m=1$ DCP frequency shift}

The $m=1$ DCP shift is caused by power flow across the cavity, arising from potential imbalances between opposing feeds or resistance inhomogeneities of the copper cavity walls. Near the cavity axis, the $m=1$ phase variations are linear gradients and lead to frequency shifts if the mean transverse positions of the atomic cloud are different on the two cavity passages, for example if the fountain is intentionally tilted~\cite{Chapelet2006,Li2010,Guena2011,Li2011}. Because power is supplied to the PTB-CSF2 cavity via a single cable to the curved waveguide, we cannot experimentally amplifiy the $m=1$ DCP shift by alternately supplying one feed or the other when the fountain is tilted~\cite{Chapelet2006,Guena2011,Li2011}. We therefore cannot use this technique to balance the feeds or to measure the tilt of the fountain to give no atomic
motion along the feed axis and no m = 1 DCP shift.

\begin{figure}
\includegraphics[width=15cm]{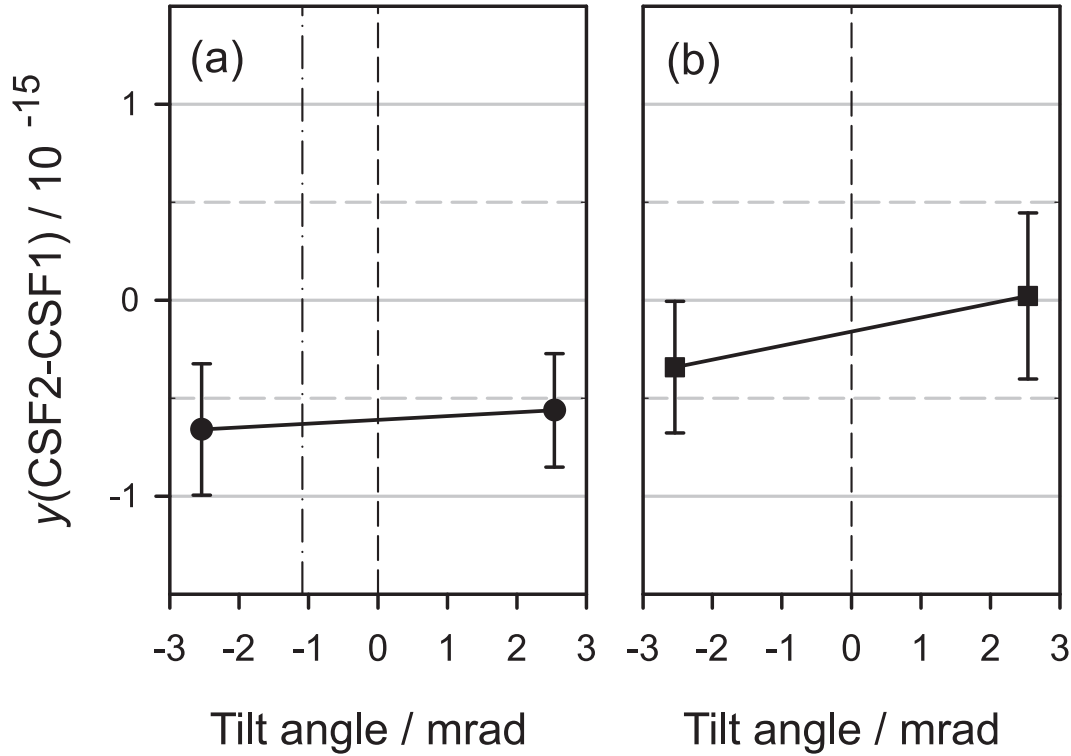}
\caption{\label{fig:m=1parperp} CSF2 frequency at optimum amplitude versus tilt, $\pm$2.54\,mrad, parallel (a) and perpendicular (b) to the feed direction. Zero tilt (dashed line) corresponds to the launch angles $\alpha_{\|}$ and $\alpha_{\perp}$ that maximize the number of detected atoms. The dash-dotted line in (a) indicates the launch angle $\alpha_{\|,0}$ at which the $m=1$ DCP contribution in the feed direction vanishes. The weighted fits (solid lines) are (a) $m_{\|}=1.9(8.7) \times 10^{-17}$mrad$^{-1}$ and (b) $m_{\perp}=7.2(10.7) \times 10^{-17}$mrad$^{-1}$.}
\end{figure}

To evaluate the $m=1$ DCP shifts, we measure the frequency change of PTB-CSF2 at optimum amplitude as we change the launch angle to $\pm$2.54\,mrad from $\alpha_{\|}$ and $\alpha_{\perp}$ that maximize the number of detected atoms, parallel and perpendicular to the feed direction (see figure~\ref{fig:m=1parperp})~\cite{Guena2011,Li2011}. Here, we again use the PTB-CSF1 fountain as the reference. Phase gradients parallel to the feeds can be caused by feed imbalances and resistance inhomogeneities, while phase gradients perpendicular to the feeds can only be from resistance inhomogeneities. The measured tilt sensitivities are $m_{\|}=1.9(8.7) \times 10^{-17}$\,mrad$^{-1}$ for parallel and $m_{\perp}=7.2(10.7) \times 10^{-17}$\,mrad$^{-1}$ for perpendicular tilts.

We need to determine the difference between the launch angle $\alpha_{\|}$ for normal operation and the launch angle $\alpha_{\|,0}$, where the $m=1$ DCP shift is zero. Note that cancelling the $m=1$ DCP shift of an offset cloud implies launching the atoms towards an even larger mean offset for the descending cloud since it in effect uniformly fills the cavity aperture. Therefore, a moderate offset implies that significantly fewer atoms are detected. For the same reason, a vertical launch gives a non-zero $m=1$ DCP shift because the ascending cloud is offset and the descending cloud is nearly centered. A Monte-Carlo simulation shows that the 1.7\,mm initial cloud offset corresponds to a launch angle difference of $\alpha_{\|}-\alpha_{\|,0}=1.09$\,mrad. Its uncertainty $\sigma_{\alpha,\|}=0.57$\,mrad includes the quadratic sum of the uncertainty of the mean position (0.27\,mrad)~\cite{Nemitz2011} and possible variations (0.5\,mrad) from long-term molasses laser-beam intensity imbalances. From $m_{\|} (\alpha_{\|,0}-\alpha_{\|})$, the $m=1$ DCP shift is $+2.1 \times 10^{-17}$ and its uncertainty is $(m_{\|}^2 \sigma_{\alpha,\|}^2+\sigma_{m,\|}^2 \sigma_{\alpha,\|}^2+\sigma_{m,\|}^2 (\alpha_{\|,0}-\alpha_{\|})^2)^{1/2}=10.8 \times 10^{-17}$.

Perpendicular to the feeds, the cloud offset is consistent with zero (see \sref{sec:Evaluation}) so that a vertical launch maximizes the number of detected atoms. This gives no $m=1$ DCP shift for atomic motion perpendicular to the feeds. The $\ket{3, 0}\rightarrow \ket{4, \pm 1}$ measurements yield an angular uncertainty of $0.33$\,mrad. Here, we again add in quadrature an 0.5\,mrad uncertainty related to possible long-term molasses laser-beam intensity imbalances. Thus, the $m=1$ DCP uncertainty for the perpendicular direction is $(m_{\perp}^2 \sigma_{\alpha,\perp }^2+\sigma_{m,\perp}^2 \sigma_{\alpha,\perp}^2)^{1/2}=7.7 \times 10^{-17}$. For both tilt directions, we get a total $m=1$ DCP shift of $2.1(13.3) \times 10^{-17}$.

\begin{figure}
\includegraphics[width=15cm]{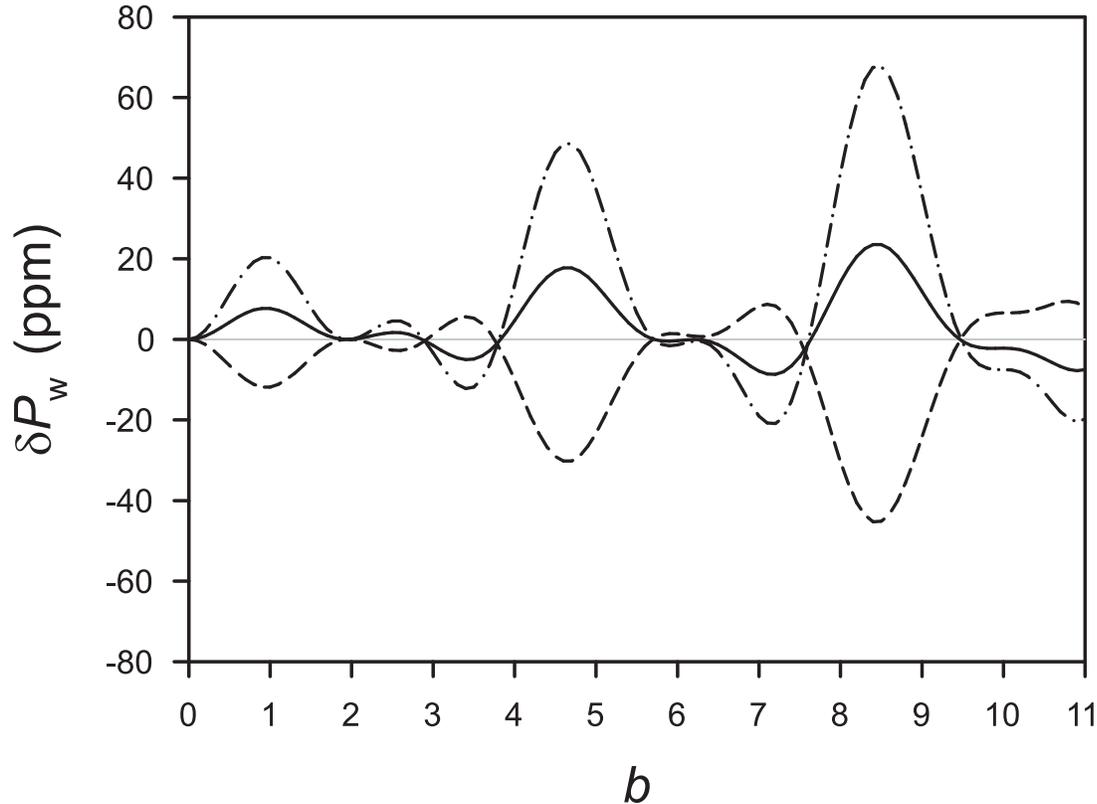}
\caption{\label{fig:m=1} Calculated $m=1$ DCP change in transition probability $\delta P_{\mbox{w}}$ of PTB-CSF2 as a function of microwave amplitude $b$ for a single weakly coupled feed. The initial cloud offset parallel to the feeds is 1.7\,mm. Solid line: tilt angle $\alpha_{\|}$ corresponding to maximum number of detected atoms. Dashed (dash-dotted) line: tilt angle $-2.54(+2.54)$\,mrad with respect to the true vertical.}
\end{figure}

It is interesting to deduce the corresponding feed imbalance from the measured parallel tilt sensitivity in figure~\ref{fig:m=1parperp}(a). In figure~\ref{fig:m=1} we show the calculated microwave amplitude dependence of the $m=1$ DCP transition probability change $\delta P_{\mbox{w}}$ of PTB-CSF2 for a single weakly coupled feed. The most likely imbalance for our cavity is an asymmetry in the machined size of the cavity couplings. Such an asymmetry supplies more power to, but also leads to greater loss from one feed. Therefore, a feed imbalance of $\epsilon$ gives $\delta P = \epsilon/2 \times \delta P_{\mbox{w}}$~\cite{Li2010}. A 10\% imbalance yields a $\delta P$ of only 5\% of $\delta P_{\mbox{w}}$ in figure~\ref{fig:m=1}. If, however, the microwave amplitude in the curved waveguide is imbalanced at the two couplings, $\delta P$ is larger by a factor $Q_0/Q$~\cite{Li2010}, where $Q_0$ is the theoretical quality factor for weak coupling, calculated to be $28600$ for PTB-CSF2, giving $Q_0/Q=17.9$. Assuming no resistance inhomogeneities, the tilt sensitivity in figure~\ref{fig:m=1parperp}(a) only constrains the couplings to be the same to 36\% and the microwave amplitudes in the curved waveguide to 2\%. We expect the balance of both is better than these estimated limits. Additionally, the measured $m=1$ DCP shift of $2.1 \times 10^{-17}$ of PTB-CSF2 at optimum amplitude corresponds to $\delta P= 0.3$\,ppm. From figure~\ref{fig:m=1}, $\delta P_{\mbox{w}}$ is $7.7$\,ppm so our measured bias suggests a suppression factor of $>25$ and in turn that $m=1$ DCP shifts are negligible in figure~\ref{fig:m=0}.

\subsection{\label{sec:m=2}$m=2$ DCP frequency shift}

The $m=2$ quadrupolar $\cos(2 \phi)$ phase variations are caused by transverse power flow from two feeds to the cavity walls. As for the $m=1$ DCP shifts, the $m=2$ DCP shift also vanishes for a horizontally symmetric cloud, launched vertically on the cavity axis and homogeneously detected. For a centered vertical launch, only the detection system can cause $m=2$ DCP shifts~\cite{Li2010}. But, if the detection laser beam propagates perpendicular to the optical axis of the imaging system and both are at 45$^\circ$ to the axis of the two cavity feeds, inhomogeneous detection will nonetheless be symmetric and produce no $m=2$ DCP shift~\cite{Li2010}. In PTB-CSF2 the orientation of the cavity feeds and the overall detection system are nearly optimal; the detection laser beam propagates at 37.5$^\circ$ from the feed axis, resulting in a suppression factor of 3.9 of the $m=2$ DCP shifts due to detection inhomogeneities. 

\begin{figure}
\includegraphics[width=15cm]{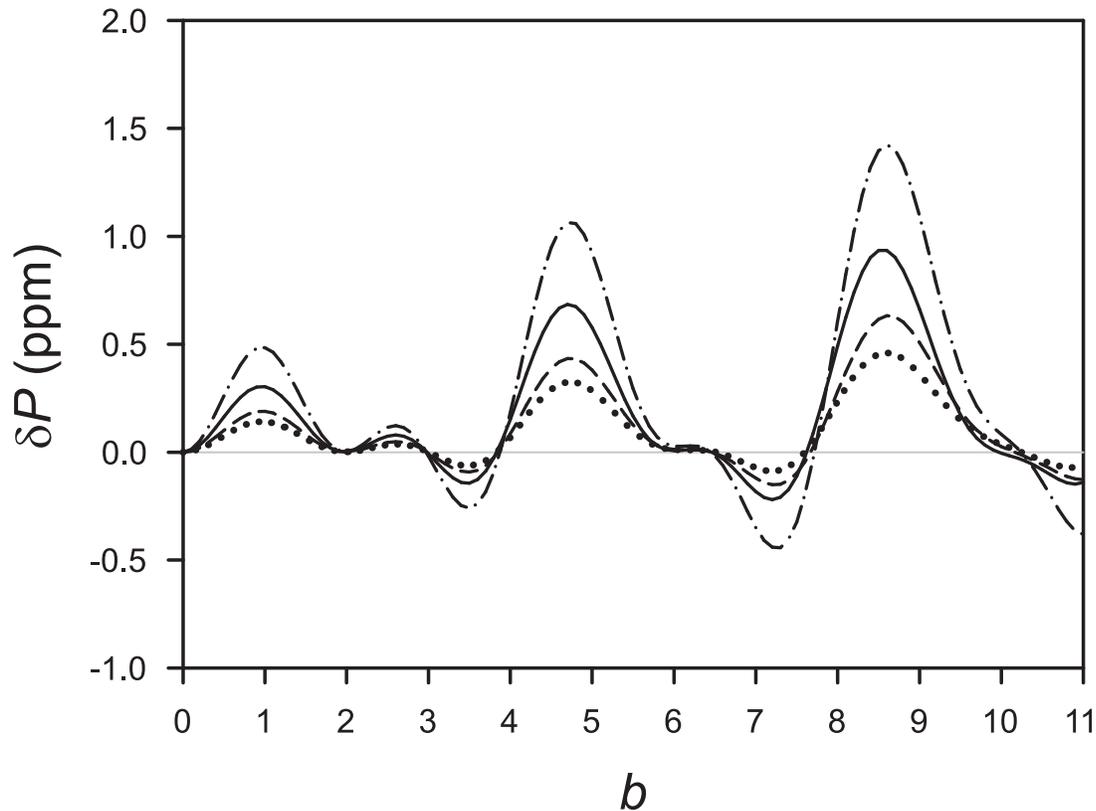}
\caption{\label{fig:m=2} Calculated $m=2$ DCP change in transition probability $\delta P$ of PTB-CSF2 as a function of microwave amplitude $b$. Solid line: the initial cloud offset in the direction of the feeds is 1.7\,mm and the horizontal launch velocity is 0.7\,mm/s. Dashed (dash-dotted) line: the initial cloud offset in the direction of the feeds is 0.8(2.6)\,mm and the horizontal launch velocity is 0.7\,mm/s. Dotted line represents zero initial cloud offset and zero horizontal launch velocity (centered vertical launch).}
\end{figure}

We evaluate the homogeneity of the detection system of PTB-CSF2 with a Monte-Carlo ray tracing simulation. We take into account the detection laser intensity profile and the geometric and optical parameters of the detection system. Because the detection laser intensity is high enough to reasonably saturate the fluorescence detection transition, the relative efficiency only varies from 100\% to 90\%. The fluorescence collection on the other hand varies much more. It has a minimum of 40\% for atoms detected around the edges of the detection zone as compared to the fluorescence collection from atoms on the fountain axis. The dotted line in figure~\ref{fig:m=2} shows the resulting calculated $\delta P$ for PTB-CSF2.

The initial cloud offset also produces an $m=2$ DCP shift. The total $m=2$ DCP shift, including the detection inhomogeneity, is the solid line in figure~\ref{fig:m=2}. At optimum amplitude, $b=0.97$, the $m=2$ DCP frequency shift is $2.1 \times 10^{-17}$. The dashed and dash-dotted lines represent transition probability changes for $\pm$50\% variations of the initial cloud offset. Considering the variations given by the dashed and dash-dotted lines as an upper limit for long-term drifts of the initial cloud offset, we get $1.4 \times 10^{-17}$ as the $m=2$ DCP uncertainty.

\subsection{\label{sec:sumDCP}Summary of DCP frequency shifts and overall uncertainty budget of PTB-CSF2}

In table~\ref{tab:table1} the individual DCP contributions $m=0,1,2$ of PTB-CSF2 are compiled and the overall correction is given, including its uncertainty. Our evaluation neglects DCP shifts from potential machining burrs~\cite{Li2010,Guena2011}. Such shifts could be avoided by using a small cloud~\cite{Li2011} or additional apertures to prevent atoms passing near the cutoff waveguide walls from being detected.

\begin{table}
\caption{\label{tab:table1} Individual DCP corrections $m=0,1,2$ of PTB-CSF2 and their uncertainties (parts in $10^{16}$).}
\begin{indented}
\lineup

\item[]\begin{tabular}{@{}lll}
\br
Effect & Correction & Uncertainty \\
\mr

DCP $m=0$                  									& $-0.018$			& 0.038\\
DCP $m=1$	(feed direction)             			& $-0.21$				& 1.08\\
DCP $m=1$	(perpendicular to feed direction)	& \m---					& 0.77\\
DCP $m=2$      						              		& $-0.21$     	& 0.14\\
\mr
total                                   		& $-0.44$		  & 1.33\\
\br
\end{tabular}
\end{indented}
\end{table}

We now include an entry in the uncertainty budget of PTB-CSF2 for the microwave lensing of the atomic wavefunctions by the microwave standing-wave~\cite{Li2011,Gibble2006}. On the fountain ascent, the interaction of the atomic wave packets with the microwave standing-wave in the clock cavity gives rise to a spatially dependent phase shift that is nearly quadratic with transverse position, thus acting as a lens on the incident atomic wave~\cite{Gibble2006}. The spatial phase variation from the dipole energy focuses one dressed state and defocuses the other during the first cavity passage. On the high frequency side of the central Ramsey fringe, the microwaves during the second cavity passage transfer the focused dressed-state to the final state of the clock transition and, on the low frequency side of the central fringe, the defocused dressed-state~\cite{Gibble2006}. As a result, the apertures in the fountain cut more of the defocused dressed-state, producing an apparent positive frequency shift of the clock transition.

The microwave lensing frequency shift is analytic if we ignore the apertures during the ascent, the microwave lens aberrations, and the detection inhomogeneities~\cite{Li2011}. This analytic result predicts a shift of $10.9 \times 10^{-17}$. Including the lower aperture of the selection cavity during the ascent reduces the shift to $7.5 \times 10^{-17}$, because it truncates the large atomic cloud far from the cavity axis where the dipole forces are large. Adding the higher order lens aberrations gives a small correction, yielding $7.1 \times 10^{-17}$. Finally, with the detection inhomogeneity included~\cite{Li2011}, the predicted microwave lensing shift is $8.3 \times 10^{-17}$ for PTB-CSF2. We take half of this correction as the uncertainty.

The updated uncertainty budget of PTB-CSF2 and the applied corrections are given in table~\ref{tab:table2}. For some corrections, i.e. quadratic Zeeman shift, blackbody radiation shift, collisional shift, typical values are given. The uncertainty due to the collisional shift is improved by a factor of $\approx 2$ compared to the first evaluation~\cite{Gerginov2010} by taking advantage of a better measurement statistics.

\begin{table}
\caption{\label{tab:table2} Updated uncertainty budget of PTB-CSF2: Frequency shift type (origin), applied frequency correction, and frequency correction uncertainty (parts in $10^{16}$).}
\begin{indented}
\lineup

\item[]\begin{tabular}{@{}lll}
\br
Frequency shift & Correction \hspace{5mm} & Uncertainty \\
\br

Quadratic Zeeman shift                  & $-1006.43$			& 0.59 \\
Blackbody radiation shift               & $\m\0164.54$    & 0.76 \\
Gravity + relativistic Doppler effect   & $-\0\085.67$    & 0.06 \\
Collisional shift                       & $\m\0\0\09.6$ 	& 3.0 \\
Cavity phase shift                      & $-\0\0\00.44$	& 1.33 \\
Microwave lensing												& $-\0\0\00.83$   & 0.42 \\
\mr
AC Stark shift (light shift)           	& $\m\0\0\00.0$   & 0.01 \\
Majorana transitions                    & $\m\0\0\00.0$   & 0.001 \\
Rabi pulling                            & $\m\0\0\00.0$   & 0.002 \\
Ramsey pulling                          & $\m\0\0\00.0$   & 0.01 \\
Electronics											        & $\m\0\0\00.0$   & 2.0 \\
Microwave leakage												& $\m\0\0\00.0$   & 1.0 \\
Background pressure                     & $\m\0\0\00.0$   & 0.5 \\
\mr
total                                   & $-\0919.2$				& 4.1 \\
\br
\end{tabular}
\end{indented}
\end{table}

The $m=0$ DCP shift accounts for our previously observed ``microwave power dependence'' so we no longer include this entry in the uncertainty budget of PTB-CSF2. We note that the currently largest uncertainty is due to collisions and is essentially given by its statistical measurement uncertainty, which will be reduced in future extended evaluations. Using rapid adiabatic passage~\cite{dossantos2002} or tuning the collision energy to cancel the shift~\cite{Szymaniec2007} could offer more profound reductions.

\section{\label{sec:conclusions} Conclusions}

We theoretically and experimentally evaluate the distributed cavity phase shifts of the PTB-CSF2 fountain clock. The DCP frequency shifts due to $m=0$ longitudinal phase gradients are measured and compared with {\it ab initio} calculations. The experimental data agrees well with the model that shows a very small correction at optimum microwave amplitude. The $m=1$ transverse phase gradients have been evaluated experimentally. At optimum microwave amplitude the gradient in the direction of the cavity feeds yields the largest DCP uncertainty contribution. The uncertainty is dominated by the statistical uncertainties of the measurements and the knowledge of the initial position and launch angle of the atomic cloud. The contribution from $m=2$ transverse phase variations was modeled, accounting for the detection inhomogeneities and the initial position and horizontal velocity of the atom cloud. We anticipate further reduction of the DCP uncertainty with better centering of the initial position of the atomic cloud. We evaluate the systematic frequency shift due to microwave lensing. Combined with a smaller uncertainty of the collision shift, the inaccuracy of PTB-CSF2 improves to $4.1 \times 10^{-16}$.

\ack

The authors gratefully acknowledge insightful discussions with R. Schr\"oder and financial support from the NSF and Penn State (R.\,L. and K.\,G.) and DFG (N.\,N.).

\end{document}